\documentclass[12pt]{article}
\usepackage{epsfig}

\begin{document}

\title{Nucleon QCD sum rules in nuclear matter including radiative
corrections }

\author{E. G. Drukarev, M. G. Ryskin, and V. A. Sadovnikova\\
Petersburg Nuclear Physics Institute\\ Gatchina, St. Petersburg
188300, Russia}

\date{}
\maketitle

\begin{abstract}
We calculate the nucleon parameters in nuclear matter using the
QCD sum rules method. The radiative corrections to the leading
operator product expansion terms are included, with the
corrections of the order $\alpha_s$ beyond the logarithmic
approximation taken into account. The density dependence of the
influence of radiative corrections on the nucleon parameters is
obtained. At saturation density the radiative corrections increase
the values of vector and scalar self-energies  by about 40~MeV,
and 30~MeV correspondingly. The results appear to be stable with
respect to possible variations of the value of $\Lambda_{QCD}$.
\end{abstract}


\section{Introduction}

In nuclear physics the nucleon parameters (self-energies) in nuclear
medium are expressed in terms of meson exchange. In QCD sum rules (SR)
approach to the problem the self-energies are expressed in terms of
exchanges by systems of weakly correlated quarks. Until now these were
just uncorrelated quarks. The correlations can be presented through the
QCD radiative corrections. In the present paper we investigate the
influence of these corrections on the values of nucleon self-energies
in symmetric nuclear matter.

The QCD sum rules (SR) method, suggested by Shifman {\em et~al.}
\cite{1} succeeded in expressing hadron characteristics in terms
of expectation values of QCD operators. This approach was
initially used for mesons. Later it was expanded by Ioffe {\em
et~al.} \cite{2,3} to description of baryons (see also \cite{3a}).
In the SR method one considers the function $\Pi(q)$ which
describes the propagation of the system with the quantum numbers
of a hadron
\begin{equation}
\Pi(q)\ =\ i\int d^4xe^{i(qx)} \langle0|Tj(x)\bar j(0)|0\rangle\,,
\end{equation}
with the local operator $j(x)$ carrying the quantum numbers of the
hadron. We shall consider the SR for proton. In this case
\begin{equation}
\Pi(q)\ =\ q_\mu\gamma^\mu\Pi^q(q^2)+I\Pi^I(q^2)
\end{equation}
with $\gamma_\mu$ and $I$ standing for the Dirac and unit $4\times4$
matrices.

The key point of the SR approach is the analysis of dispersion
relations
\begin{equation}
\Pi^i(q^2)\ =\ \frac1\pi \int\frac{\mbox{Im }\Pi^i(k^2)}{k^2-q^2}\,dk^2
\end{equation}
for the functions $\Pi^i(q^2)$. These equations are considered at
large values of $|q^2|$, while $q^2<0$. The left-hand sides (LHS)
of Eq.~(3) are presented as power series of $q^{-2}$, with the QCD
condensates being the coefficients of the expansion. This
presentation is know as the operator product expansion (OPE)
\cite{4}. The spectral densities Im$\,\Pi^i(k^2)$ on the
right-hand side (RHS) of Eq.~(3) are usually approximated by the
``pole+continuum" model
\begin{equation}
\mbox{Im }\Pi^i(q^2)\ =\ \lambda^2_N\,\delta(k^2-m^2)+\frac1\pi\,\theta
(k^2-W^2)\,\mbox{Im }\Pi^{i\rm\,OPE}(k^2)\,.
\end{equation}
Thus the position of the lowest pole $m$, its residue
$\lambda^2_N$ and the effective continuum threshold $W^2$ are the
unknowns of the SR equations in vacuum. Usually the Borel
transform of both sides of Eq.~(3) is carried out.

Several lowest order OPE terms have been found in \cite{2,3}. These
were the contribution of free three-quark loop and the terms containing
the vacuum condensates $\langle0|\bar qq|0\rangle$,
$\langle0|\frac{\alpha_s}\pi G_{\mu\nu}^a G^a_{\mu\nu}|0\rangle$ and
$\langle0|\bar qq\bar qq|0\rangle$, with $q$ and $G^a_{\mu\nu}$ the
quark operators and the gluon field tensor.

Later this approach was expanded for description of nucleons in
nuclear matter \cite{5,6,6a}(see also \cite{7} and references
therein). In this case there are three structure of the
polarization operator
\begin{equation}
\Pi_m(q,p)=i\int d^4xe^{i(qx)}\langle M|\bar j(x)\bar
j(0)|M\rangle
=q_\mu\gamma^\mu\Pi_m^q(q,p)+p_\mu\gamma^\mu\Pi_m^p(q,p)+I\Pi_m^I(q,p),
\end{equation}
with $|M\rangle$ and $p$ -- the vector of state of the matter and
its four-momentum; $p=(m,0)$ in the rest frame of the matter. The
dispersion relations for the functions $\Pi^i(q,p)=\Pi^i(q^2,s)$
at fixed value of $s=(p+q)^2$
\begin{equation}
\Pi^i_m(q^2,s)\ =\ \frac1\pi\int\frac{\mbox{Im
}\Pi^i_m(k^2,s)}{k^2-q^2}\,dk^2
\end{equation}
are considered now instead of Eq.~(3). The same approximations as
in vacuum, i.e. the OPE and "pole+continuum" model
were used for the LHS and RHS of the SR. The LHS
contains now contributions of two types. There are expectation
values of the same operators as in the case of vacuum, averaged
over the ground state of nuclear matter $(\langle M|\bar
qq|M\rangle$, etc.). There are also the QCD condensates which
vanish in vacuum. In the leading OPE terms these are the vector
condensate $\langle M|\bar q(0)\gamma_\mu q(0)|M\rangle$ and the
expectation value $\langle M|\bar q(0)\gamma_\mu D_\nu
q(0)|M\rangle$ caused by the nonlocal vector operator $\langle
M|\bar q(0)\gamma_\mu q(x)|M\rangle$.

The in-medium characteristics of the nucleons, i.e. the vector
self-energy $\sum_v$ and the Dirac effective mass $m^*$ are
unknowns of the SR equations in nuclear matter. Two other unknowns
are the residue at the nucleon pole and the continuum threshold,
which obtain new values $\lambda^2_m$ and $W^2_m$. The SR approach
\cite{5,6,6a,7} provided reasonable values of
$\Sigma_v\approx200\,$MeV, $m^*-m\approx-300\,$MeV, which are
consistent with the results obtained by nuclear physics methods.
Similar results have been obtained in other modifications of the
SR approach \cite{8,9}.

Interactions between the quarks in polarization operators (1) and
(5) manifest themselves in radiative corrections, which contain
the powers of the QCD coupling constant $\alpha_s$. In the
asymptotics $q^2\to-\infty$ the terms in which $\alpha_s$ is
enhanced by the ``large logarithm" $\ln(q^2)$ are the most
important ones. The corrections $(\alpha_s\ln q^2)^n$ have been
obtained in all orders for the leading OPE terms \cite{10} and
included into calculations carried out in \cite{2,3,5,6,7}.
The SR are considered at finite values of the Borel mass
$M^2$, i.e.
\begin{equation}
0.8\mbox{ GeV}^2\ <\ M^2\ <\  1.4\mbox{ GeV}^2,
\end{equation}
where $\alpha_s$ is small enough for perturbative treatment of the
radiative corrections. Perturbative calculation of such
corrections in vacuum beyond the logarithmic approximation
\cite{11,13} provided a numerically large coefficient of the
lowest radiative correction to the leading OPE terms. Analysis of
the role of radiative corrections in vacuum Borel transformed SR
method which includes also corrections to the four-quark
condensate \cite{13} was carried out in \cite{14}. It was shown
that at least for the standard current
\cite{15}
$$j(x)= \varepsilon_{abc}[u_a^T(x)C\gamma_{\mu}u_b(x)]\gamma_5\gamma^{\mu}d_c(x)$$
these
corrections modify mainly the value of the nucleon residue
$\lambda^2_N$, while that of the nucleon mass suffers minor
changes. In the present paper we are also using  this current.

Recent calculations \cite{16} demonstrated that the lowest order
radiative corrections to the leading OPE terms of the in-medium
polarization operator (5) contain the coupling constant $\alpha_s$
multiplied by large coefficients, which are  7/2 in $\Pi^q_m$
structure and 15/4 in $\Pi^p_m$ structure. Since
$\alpha_s/\pi\approx0.15$ at the values of the Borel mass
$M^2\approx1\rm\,GeV^2$, new analysis of the sum rules with
inclusion of radiative corrections is needed.

We present the polarization operator $\Pi_m$ determined by Eq.~(5) as
\begin{equation}
\Pi_m(q^2,s)\ =\ \Pi(q^2)+\Pi_\rho(q^2,s)\,,
\end{equation}
with the vacuum term $\Pi$ determined by Eq.~(1) while
$\Pi_\rho(q^2,s)$ describes interaction with the nuclear matter.
We include radiative corrections to the vacuum term up to the
contributions $\sim q^{-2}$ of OPE. We include also  radiative
corrections to the leading OPE terms provided by interaction with
nuclear matter. We compare the results provided by inclusion of
radiative corrections in several ways.

It is well known that the sum of the terms $(\alpha_s\ln q^2)^n$ can be
presented in terms of the function
\begin{equation}
L(q^2)\ =\ \frac{\ln q^2/\Lambda^2}{\ln\mu^2/\Lambda^2}
\end{equation}
with $\Lambda=\Lambda_{\rm QCD}$while $\mu=500\,$MeV is the
normalization point of the characteristic involved. The sum of
corrections $(\alpha_s\ln q^2)^n$ to a certain term multiplies it
by a factor $L^{-\gamma}$, with $\gamma$ reflecting the``anomalous
dimension" of corresponding operator -- see, $i.e.$ \cite{17}.

In the present paper we consider the Borel transformed sum rules
for the nucleon in nuclear matter, considering the dispersion
relations for the operator $\Pi_\rho$ defined by Eq.~(8),
employing the current $j(x)$ suggested in \cite{15}. We follow our
paper \cite{7} adding the radiative corrections to the analysis
carried out there. We include the logarithmic corrections and the
correction of the order $\alpha_s$ beyond the leading logarithmic
approximation (LLA) to the condensates of dimension $d=3$, which
are the vector and scalar condensates. We include also the LLA
corrections to the main condensates of dimension $d=4$, which are
due to nonlocal structure of the vector condensate.

We demonstrate that while the LLA corrections provide substantial
changes of the values of vector and scalar nucleon self-energies
$\Sigma_v$ and $m^*-m$, these changes are to large extent
compensated by the terms $\sim\alpha_s$, which are beyond the LLA.
For example, at saturation value of nucleon density this
composition of the radiative corrections adds 40\,MeV and 30\,MeV
to $\Sigma_v$ and $m^*-m$ correspondingly. These corrections
effect mostly the values of the vector self-energy and of the
nucleon residue. The values of $m^*-m$ and of the continuum
threshold suffer minor changes.

Our main numerical results are obtained for $\Lambda_{\rm
QCD}=150\,$MeV corresponding to $\alpha_s(1\rm\,GeV^2)=0.37$ in
one-loop approximation. We used these values in our previous
papers on the subject. Somewhat larger values of $\Lambda_{\rm
QCD}$ and $\alpha_s(1\rm\,GeV^2)$ are often used nowadays -- see,
e.g. \cite{3a}. The possible modifications affect mostly the value
of the nucleon residue, while the results for the self energies
appear to be stable within several percent. We recall the results
of inclusion of the radiative corrections to vacuum sum rules in
Sect.~II adding also some new data. The sum rules in nuclear
matter with radiative corrections are considered in Sect.~III. The
results are discussed in Sect.~IV.

\section{Radiative correction in vacuum}

Following \cite{14} we present the lowest OPE terms of the
operators $\Pi^q(q^2)$ and $\Pi^I(q^2)$ as
\begin{equation}
\Pi^q=A_0+A_4+A_6+A_8\,; \quad \Pi^I=B_3+B_7+B_9\,,
\end{equation}
with lower indices showing the dimensions of the condensates, $A_0$ is
the contribution of the free 3-quark loop. Perturbative inclusion of
the lowest order $\alpha_s$ corrections provides
\begin{eqnarray}
A_0 &=& -\frac1{64\pi^4}\,Q^4\ln\frac{Q^2}{\mu^2}\left(1+\frac{71}{12}
\,\frac{\alpha_s}\pi
-\frac12\,\frac{\alpha_s}\pi\ln\frac{Q^2}{\mu^2}\right);
\nonumber\\
A_6 &=& \frac23\,\frac{\langle0|\bar qq\bar qq|0\rangle}{Q^2}\left(
1-\frac56\,\frac{\alpha_s}\pi-\frac13\,\frac{\alpha_s}\pi
\ln\frac{Q^2}{\mu^2}\right);
\nonumber\\
B_3&=& \frac{-\langle0|\bar qq|0\rangle}{4\pi^2}\,Q^2
\ln\frac{Q^2}{\mu^2}\left(1+\frac{3\alpha_s}{2\pi}\right),
\end{eqnarray}
with $Q^2=-q^2>0$. Radiative corrections to the terms $A_4$ and $B_7$
which are numerically small \cite{2,3}, and to the terms $A_8$ and
$B_9$, which are known with poor accuracy are not included.

The Borel transformed nucleon sum rules \cite{2,3} can be written as
\begin{equation}
{\cal L}^q(M^2,W^2)={\cal R}^q(M^2)\,, \quad {\cal
L}^I(M^2,W^2)={\cal R}^I(M^2)\,,
\end{equation}
with
\begin{equation}
{\cal R}^q(M^2)=\lambda^2e^{-m^2/M^2}, \quad {\cal
R}^I(M^2)=\lambda^2me^{-m^2/M^2}
\end{equation}
describing the contributions of nucleon pole
$(\lambda^2=32\pi^4\lambda^2_N)$, while
\begin{equation}
{\cal L}^q=\tilde A_0+\tilde A_4+\tilde A_6+\tilde A_8\,, \quad
{\cal L}^I=\tilde B_3+\tilde B_7+\tilde B_9\,.
\end{equation}
Here $\tilde A_i(\tilde B_i)$ denote the Borel transforms of the terms
$A_i(B_i)$ on the RHS of Eq.~(10), multiplied also by $32\pi^4$.

Explicit expression for $\tilde A_i$ and $\tilde B_i$ are given in
\cite{14}. Here we focus on the terms which contain the radiative
corrections. Including the contributions $(\alpha_s\ln q^2)^n$ exactly
and treating the corrections $\sim\alpha_s$ beyond the logarithms
perturbatively, we can write
\begin{eqnarray}
&& \tilde A_0=\frac{M^6E_2(W^2/M^2)r_0}{\tilde L^{4/9}}\,; \quad
\tilde B_3 =\frac{2a(M^2)M^4E_1(W^2/M^2)r_3}{\tilde L^{4/9}}\,;
\nonumber\\
&& \tilde A_6\ =\ \frac43\,\frac{a_4(M^2)}{\tilde L^{4/9}}\, r_6\, .
\end{eqnarray}
Here $E_1(x)=1-(1+x)e^{-x}$; $E_2(x)=1-(1+x+x^2/2)e^{-x}$. The function
\begin{equation}
\tilde L(M^2)\ =\ \frac{\ln M^2/\Lambda^2}{\ln\mu^2/\Lambda^2}
\end{equation}
comes from logarithmic corrections to the current $j(x)$ and to
condensates. Also in Eq.~(15)
\begin{equation}
a=-(2\pi)^2\langle0|\bar q q|0\rangle\,; \quad
a_4=(2\pi)^4\langle0| \bar q q\bar q q|0\rangle\,.
\end{equation}
For $i=0,3$ the factors
\begin{equation}
r_i=1+\frac{\alpha_s}\pi\,c_i\,; \quad c_0=\frac{53}{12}\,, \quad
c_3=\frac32\,
\end{equation}
include the radiative corrections of the order $\alpha_s$ beyond the LLA
\cite{13}.

The situation with the contribution $\tilde A_6$ is a bit more complicated.
The four-quark condensate is presented usually under factorization
assumption \cite{1}
\begin{equation}
\langle0|\bar q q\bar q q|0\rangle\ =\ (\langle0|\bar q q|0\rangle)^2,
\quad a_4=a^2.
\end{equation}
However, one should clarify the point, where Eq.~(19) is
postulated. Assuming, following \cite{3a} that the factorization
$a_4(k^2)=a^2(k^2)$ to be true at certain $k^2=M_0^2$, with
$M_0^2$ belonging to the interval determined by Eq.(7) and
employing $a(M_0^2)=a(\mu^2)\tilde L^{4/9}$ one finds in the LLA
\begin{equation}
a_4(M_0^2)=a^2(\mu^2)\tilde L(M_0^2)^{8/9}.
\end{equation}
 On the other hand
$M^2$ dependence of the condensate $a_4$ in framework of the
factorization hypothesis was found in \cite{13}
\begin{equation}
a_4(M^2)=a_4(M_0^2)\left(1-\frac{1}{3}\frac{\alpha_s}{\pi}\ln\frac{M^2}{M_0^2}\right).
\end{equation}
Of course, $\ln \frac{M^2}{M^2_0}$ is not a "large logarithm" and hence is included perturbatively.
Thus we can write
\begin{equation}
\tilde B_3=2a(\mu^2)M^4E_1(W^2/M^2)r_3\,, \quad \tilde
A_6=\frac43\,a^2(\mu^2)\tilde L(M_0^2)^{4/9}r_6(M^2,W^2).
\end{equation}
Here
\begin{equation}
r_6(M^2, W^2)=1-\frac{\alpha_s}{3\pi}\left(\frac{5}{2}+\ln
\frac{W^2}{M_0^2}+{\cal E}\left(-\frac{W^2}{M^2}\right)\right),
\end{equation}
with
$${\cal E}(x)=\sum_{n=1}\frac{x^n}{n\cdot n!}$$
includes the lowest correction beyond the LLA.

In Eqs. (15) and (22) we assumed that the corrections of the order
$(\alpha_s\ln M^2)^n$ and of the order $\alpha_s$ compose
independent factors. This can be justified by the fact that the
former terms come from integration over momenta $\mu^2\ll k^2\ll
M^2$, while  the latter come from $k^2\sim M^2$.


In Table I we present the values of characteristics $m,\lambda^2$
and $W^2$ obtained with the radiative corrections being included
in various ways. We show the results with the radiative
corrections totally neglected (i.e. all $\tilde L=1$, $r_i=1$),
the results in LLA based on Eq.(22) with $\tilde L$ determined by
Eq.(16), while $r_i=1$. We include the corrections beyond LLA,
with $r_i$ determined by Eqs.(18) and (23). We show also the
results for all perturbative inclusion of radiative corrections
(PRIC)- see Eq.(11). We assume the factorization point for the
four-quark condensate (Eq.(19)) to be $M_0^2=1~$GeV$^2$. We
present also the results corresponding to factorization point
$M_0=\mu=0.5~$GeV obtained earlier in \cite{14}.


\section{Radiative corrections in nuclear matter}
\subsection*{1. Sum rules without radiative corrections}

Following \cite{5,6,7} we consider the SR for
\begin{equation}
\Pi^q_\rho=A_{3\rho}+A_{4\rho}+A_{6\rho}\,; \quad
\Pi^I_\rho=B_{3\rho}+B_{6\rho}\,; \quad
\Pi^p_\rho=P_{3\rho}+P_{4\rho}+P_{6\rho}\,.
\end{equation}
The Borel transformed SR are
\begin{eqnarray}
&& {\cal L}^q_\rho(M^2,W^2_m)=\tilde A_{3\rho}+\tilde A_{4\rho}
+\tilde A_{6\rho}=\Lambda_m(M^2)-\Lambda(M^2);
\\
&& {\cal L}^I_\rho(M^2,W^2_m)=\tilde B_{3\rho}+\tilde B_{6\rho}
=m^*\Lambda_m(M^2)-m\Lambda(M^2);
\\
&& {\cal L}^p_\rho(M^2,W^2_m)=\tilde P_{3\rho}+\tilde P_{4\rho}
+\tilde P_{6\rho}=-\Sigma_v\Lambda_m(M^2),
\end{eqnarray}
with $\Lambda_m(M^2)=\lambda^2_m e^{-m^2_m/M^2}$,
$\Lambda(M^2)=\lambda^2e^{-m^2/M^2}$; $\lambda^2_m$ and $W^2_m$
are the effective value of the residue and of the continuum
threshold in nuclear matter; $m^2_m$ can be presented in terms of
$\Sigma_v$ and $m^*$ \cite{7}.

The lowest dimension terms can be written  as \cite{7}
\begin{eqnarray}
A_{3\rho} &=& \frac{\langle M|\sum\limits_i \bar q^i\gamma_0 q^i|M\rangle}{
6\pi^2 m}\,(pq)\ln Q^2\,;
\nonumber\\
P_{3\rho} &=& -\frac{\langle M|\sum\limits_i \bar q^i\gamma_0 q^i|M\rangle
Q^2\ln Q^2}{3\pi^2 m}\,,
\end{eqnarray}
($i=u,d$)\\
and
\begin{equation}
B_{3\rho}=\frac{-\langle N|\sum\limits_i\bar q^iq^i|N\rangle\rho}{8\pi^2}
Q^2\ln Q^2, \quad \langle N|\sum\bar q^iq^i|N\rangle\approx8.
\end{equation}
Since the SR are considered at $s=(p+q)^2={\rm const}=4m^2$
\cite{5,6,6a,7}, we must put $(pq)=(s-m^2-q^2)/2$ in Eq.~(28).
From Eqs. (28), (29) we obtain \cite{7}
\begin{eqnarray}
&& \tilde A_{3\rho} =-8\pi^2\langle M|\sum\limits_i \bar q^i\gamma_0 q^i|M\rangle
\frac{(s-m^2)E_0(M^2)-M^2E_1(M^2)}{3m};
\nonumber\\
&&\tilde B_{3\rho}=-4\pi^2\langle N|\sum\limits_i\bar q^iq^i|N\rangle M^4E_1(M^2)\rho; \nonumber\\
&&\tilde P_{3\rho}= -\frac{32\pi^2}{3}\langle M|\sum\limits_i \bar q^i\gamma_0 q^i|M\rangle
M^4E_1(M^2).
\end{eqnarray}

The contributions of the fourth dimension $A_{4\rho}$ and $P_{4\rho}$
come mainly from the nonlocality of the vector condensate
$\langle0|\sum\limits_i \bar q^i(0)\gamma_0D_\mu q^i(x=0)|0\rangle$ (there
is also a small contribution $A^g_{4\rho}$ caused by the in-medium
gluon condensate -- see \cite{7}). Such contributions can be expressed
in terms of the nucleon structure functions \cite{5,6,7}
\begin{eqnarray}
&& A_{4\rho}\ =\ A^g_{4\rho}+A^v_{4\rho}\, ;
\nonumber\\
&& A^v_{4\rho}=\frac{m}
{6\pi^2}\ln\frac{Q^2}{\Lambda^2_c}\int\limits^1_0
d\alpha\Big(-\alpha \eta_a(\alpha)+9\eta_b(\alpha)\Big)\rho
\nonumber\\
&&+\ \frac1{6\pi^2m}\int\limits^1_0 d\alpha\ln\frac{Q'^{2}}{Q^2}
\Big((pq')\eta_a(\alpha)+9m^2\eta_b(\alpha)\Big)\rho
\end{eqnarray}
  and
\begin{eqnarray}
&& P^v_{4\rho}=-\frac{1}{6\pi^2m}\ln\frac{Q^2}{\Lambda^2_c}
\int\limits^1_0
d\alpha\Big[(5\alpha(pq')+2\alpha^2m^2)\eta_a(\alpha) +9m^2\alpha
\eta_b(\alpha)\Big]\rho+
\nonumber\\
&&+\ \frac1{6\pi^2m}\int\limits^1_0 d\alpha\ln\frac{Q'^{2}}{Q^2}
\Big[(-\alpha(pq')-2Q'^{2})\eta_a(\alpha)-9m^2\alpha
\eta_b(\alpha)\Big]\rho.
\end{eqnarray}
Here $q'=q-p\alpha$, $Q^2=-q^2$, $Q'^{2}=-q'^{2}$, the functions
$\eta_{a,b}(\alpha)$ are defined following \cite{6a,7}
\begin{eqnarray*}
&& \hspace*{-0.3cm}
\langle M|\bar q(0)\gamma_\mu q(x)|M\rangle\ =
\\
&&=\left(\frac{p_\mu}m\int\limits^1_0 d\alpha\,e^{-i\alpha(px)}
\eta_a(\alpha)+ix_\mu m\int\limits^1_0 d\alpha\,e^{-i\alpha(px)}
\eta_b(\alpha)\right)\rho,
\end{eqnarray*}
with $\eta_a(\alpha)$ the standard deep inelastic nucleon structure
function, the moments of the function $\eta_b(\alpha)$ can be expressed in
terms of those of the function $\eta_a(\alpha)$ \cite{6a}. The cut off
$\Lambda_c$ will be eliminated by the Borel transform.

For the Borel transforms of the left hand sides of Eqs. (31) and (32) (multiplied
by $32\pi^4$) we find
\begin{equation}
\tilde A^v_{4\rho}(M^2)=u^q(M^2)+h^q(M^2);\ \tilde P_{4\rho}(M^2)
=u^p(M^2)+h^p(M^2),
\end{equation}
with the contributions $u^q$ and $u^p$ coming from the first terms
on the right hand sides of Eqs. (31) and (32), while $h^q(M^2)$ and $h^p(M^2)$
come from the second terms. Expressions for $u^q$ and $u^p$ are
presented in \cite{7}, while $h^q$ and $h^p$ require a special
treatment. Due to the terms $\ln Q'^{2}/Q^2$ they have finite
cuts, corresponding to the singularities in the $u$ channel of the
interaction between the baryon current and the quarks belonging to
the nucleon of matter, i.e. by the exchange terms. In paper
\cite{7} these terms have been neglected. This was justified by
claiming for the description of the nucleon in the Hartree
approximation.

Here we include the terms $h^q$ and $h^p$.
Performing the
the Borel transform of
$$
\ln\frac{Q'^{2}}{Q^2}=\ln\frac{(1+\alpha)(Q^2+X^2(\alpha))}{Q^2}\,,
$$
with $X^2(\alpha)=\frac\alpha{1+\alpha}(s-m^2-m^2\alpha)$
and using the numerical values
of the moments of the structure functions $\eta_a$ \cite{18}, we find the
coefficients of $M^{-2}$ expansion of the terms containing $\eta_a$ to be of
the same order of magnitude.
Thus the OPE in powers of $M^{-2}$
series for these terms exhibits a poor convergence.
Hence, following \cite{6a}, we employ the
explicit expressions
\begin{eqnarray}
h^q(M^2) &=& \frac{8\pi^2}{3m}\int\limits^1_0d\alpha\Bigg[\Big(
(s-m^2-2m^2\alpha)G_0(M^2,\alpha)-G_1(M^2,\alpha)\Big)
\nonumber\\
&&\times\ \eta_a(\alpha)- 18m^2\left(M^2-(s-m^2)\alpha)\right)
\eta_b (\alpha)\Bigg],
\\
h^p(M^2)&=&\frac{8\pi^2}{3m}\int\limits^1_0 d\alpha\bigg[\Big((-5(s-m^2)
\alpha+6m^2\alpha^2)G_0(M^2,\alpha)+
\nonumber\\
&& +\ (4+5\alpha)G_1(M^2,\alpha)\Big)
\eta_a(\alpha)-18m^2\alpha\eta_b(\alpha)\bigg].
\end{eqnarray}
Here $G_0(M^2,\alpha)=M^2(1-e^{-X^2(\alpha)/M^2})$;
$G_1(M^2,\alpha)=M^4\left(1-(1+X^2(\alpha)/M^2)e^{-X^2(\alpha)/M^2}
\right)$. We included only the lowest moments of the function
$\eta_b(\alpha)$, which are related to the moments of the function
$\eta_a(\alpha)$ by the equations, presented in \cite{6a}. The values
of these moments enable to expect the convergence of the latter
expansion.

The contributions $A_{6\rho}$, $B_{6\rho}$ and $P_{6\rho}$ are
determined by the 4-quark condensate. Here the calculations require
certain model assumptions. We shall use the expressions presented in
\cite{7}, based on the calculations of the four-quark condensates
\cite{19} in framework of the Perturbative Chiral Quark Model suggested
in \cite{20}. Further development of this model is described in
\cite{20a}.

\subsection*{2. Inclusion of radiative corrections}

Now we include the radiative corrections. If the LLA corrections
$(\alpha_s\ln q^2)^n$ are included exactly, while the terms beyond the
LLA are taken into account in the lowest order of
perturbation theory,
each term of Eq.(30),obtains a factor
\begin{equation}
\xi^x=\frac{r^x}{\tilde L^{\gamma^x}}
\end{equation}
with $x=q,I,x$, and
$r^x=1+\frac{\alpha_s}\pi c^x$.

Following \cite{10,3,6} we put
$\gamma^q=\gamma^p=\gamma^q=4/9$;
$\gamma^I=0$.
In the LLA
$c^i_{nx}=0$.
Beyond the LLA we employ
the recently obtained in-medium parameters $c^q=7/2$ and
$c^p=15/4$ \cite{16} for corrections to the local vector
condensate, while $c^I=3/2$ \cite{13}.

Considering the terms of the higher dimension, we include the
factor $\tilde L^{-4/9}$ corresponding to the anomalous dimension of the proton current.
Note also that the nucleon structure functions
\cite{18} employed in our paper reproduce their moments with the
proper anomalous dimensions.
Corrections to contributions of the nonlocal vector condensate beyond
the logarithmic approximation require calculation of additional set of
the Feynman diagrams. As it stands now such corrections are not known.


We do not include the radiative corrections of
the contributions $A_{6\rho}$, $B_{6\rho}$ and $P_{6\rho}$ to the
four-quark condensates, provided by the nucleons of the matter. This is
because these contributions are obtained in framework of a certain
model \cite{20}, whose accuracy cannot be estimated.

Thus we include the radiative corrections BLLA to the leading OPE
contributions containing condensates of dimension $d=3$. We
include the LLA corrections to the terms containing condensates of
dimension $d=4$. We do not include corrections to the terms with
the condensates of higher dimension. Such approach is possible
since the leading OPE terms indeed provide numerically largest
contributions. For example, considering Eq.(27) for $\Sigma_v$ and
employing the vacuum values of parameters $\lambda_m^2$ and
$W^2_m$, presented in Table 1 we can see that at the saturation
value of nucleon density $\rho_0$ the local vector condensate
contributes approximately $330$ MeV to $\Sigma_v$, while nonlocal
corrections and the four-quark condensate subtract about $50$ MeV
and $70$ MeV from this value.

We compare several sets of results, corresponding to the cases, considered for vacuum in Sec.II.

\subsection*{3. Numerical results}

Now we present the results of solutions of the SR equations.
Following our previous works we put $\alpha_s(\rm 1\,GeV^2)=0.37$
which is consistent with the present data \cite{21}. At the
saturation density $\rho_0$ we find the vector self-energy
$\Sigma_v=230\,$MeV, while $m^*-m=-330\,$MeV, if the radiative
corrections are totally neglected.


Inclusion of the leading logarithmic corrections diminishes $\Sigma_v$ and
$m^*-m$ by $70$ MeV and $50$ MeV correspondingly.
Inclusion of radiative corrections beyond the logarithms makes the
values of $\Sigma_v$ and $m^*-m$ closer to
those, obtained with total neglection of the radiative corrections. The
data at saturation density are presented in Table~II. Comparing the results of BLLA calculation
and those obtained with perturbative inclusion of all $\alpha_s$ corrections, we see
that they differ by about $20 \%$. Thus it is important to include  corrections
$(\alpha_s\ln q^2)^n$ in all orders.
Density dependence
of nucleon parameters is shown in Figs.~1, 2.


Somewhat larger values of $\alpha_s(1\rm\,GeV^2)$ are often used
in nowadays calculations. The authors of \cite{16} employed
$\alpha_s=0.47$ corresponding to $\Lambda_{\rm QCD}=230\,$MeV in
one-loop approximation, while the value
$\alpha_s(1\rm\,GeV^2)=0.55$ corresponding to $\Lambda_{\rm
QCD}=280\,$MeV was used in \cite{3a}. We carried out calculations
with these sets of parameters as well. Assuming
$\alpha_s(1\rm\,GeV^2)=0.55$ we found very small changes of values
of the vector self-energy $\Sigma_v \approx 260$ MeV and of the
scalar one $m^*-m \approx -290$ MeV. The residue $\lambda^2$
suffered largest change, reaching the value $1.8~$GeV$^6$. Recall
that in the case of vacuum sum rules variation of the value of
$\alpha_s$ also affected mostly the value of $\lambda^2$
\cite{14}. Dependence of the results on the value of
$\Lambda_{QCD}$ are shown in Table III.

\section{Summary}

We calculated nucleon self-energies $\Sigma_v$ and $m^*-m$ in nuclear
matter in framework of the QCD sum rules approach, including the
radiative corrections to the leading OPE terms. We demonstrated that
there is a large compensation between the logarithmic corrections
$(\alpha_s\ln q^2)^n$ and the corrections of the order $\alpha_s$
beyond the logarithmic approximation. We showed that at saturation
value of nuclear density simultaneous inclusion of these corrections
diminishes  the values of the vector self-energy $\Sigma_v$ by about 40~MeV and of the
scalar self- energy by about 39~MeV. The threshold value suffers minor changes.
The radiative corrections affect mostly the value of the nucleon residue $\lambda^2_m$.

Result for the saturation density are presented in Table~II. Density
dependence of the parameters is shown in Figs. 1, 2.

We analyzed also stability of the results with respect to variation
of $\Lambda_{\rm QCD}$ and $\alpha_s(\rm1\,GeV^2)$ since other sets of values are often
used in literature. We found the changes to affect mostly the value of the nucleon residue. The vector
self energy $\Sigma_v$ and the scalar self energy $m^*-m$ may vary only by
percents at the saturation density. The results are given in Table III.

\newpage
\begin{table}
\caption{Nucleon parameters in vacuum. Line 1-all radiative
corrections are neglected. Line 2 -radiative corrections are
included in the leading logarithmic approximation (LLA). Line 3 -
corrections $\sim\alpha_s$ are included beyond the LLA (BLLA).
Lines 4 - results for perturbative inclusion of radiative
corrections (PIRC~1)  $\alpha_s$ and $\sim\alpha_s\ln q^2$ with
factorization (19) assumed at $M_0^2=1~$GeV$^2$. Line 5-(PIRC-2)
with factorization assumed at $M_0=0.5$~GeV}
\begin{center}
\begin{tabular}{|l|c|c|c|}
\hline
& $m$, GeV & $\lambda^2,\rm GeV^6$ & $W^2,\rm GeV^2$ \\
\hline
Rad. corrections &&&\\
are neglected                        & 0.94 & 1.84 & 2.03\\
LLA  & 0.96 & 2.13 & 2.37\\

BLLA & 0.91 & 2.03 & 1.96\\

PIRC~1 & 0.92 & 1.94 & 1.85\\

PIRC~2 & 0.95 & 2.04 & 1.91\\

 \hline
\end{tabular}
\end{center}
\end{table}

\begin{table}
\caption{Nucleon parameters at the phenomenological saturation
value of nucleon density. Notations are the same as in Table.~I}
\begin{center}
\begin{tabular}{|c|c|c|c|c|} \hline
 &  $\Sigma_v$, MeV & $m^*-m$, MeV & $\lambda^2_m\rm\ ,GeV^6$ &
$W^2_m,\rm\,$GeV$^2$ \\
\hline
Rad. corrections &  &  &  & \\
are neglected & 229 & --329 & 1.10 & 1.72\\

LLA & 160 & --380& 1.10 & 1.77\\

BLLA & 271 & --300 & 1.41 & 1.75\\

PIRC~1 & 334& --251& 1.47&1.80 \\

PIRC~2 & 347& --249& 1.48& 1.81\\
\hline
\end{tabular} \end{center}
\end{table}

\newpage
\begin{table}
\caption{Dependence of the nucleon parameters on the values of
$\Lambda_{QCD}$ and $\alpha_s$ at the phenomenological saturation
value of nucleon density with radiative corrections included
beyond the Leading logarithmic Approximation (BLLA). The results
are presented for $\Lambda_{QCD}=0.23 $ GeV,
$\alpha_s(1$GeV$^2)=0.47$ and $\Lambda_{QCD}=0.23$ GeV,
$\alpha_s(1 $GeV$^2)=0.47$. The corresponding values of vacuum
parameters are given in brackets.}
\begin{center}
\begin{tabular}{|c|c|c|c|c|} \hline
$\alpha_s(1$~GeV$^2) $&  $\Sigma_v$, MeV & $m^*-m$, MeV &
$\lambda^2_m\rm\ ,GeV^6$ &
$W^2_m,\rm\,$GeV$^2$ \\
\hline
0.47 & 269 & -291 ($m=0.93$ GeV)& 1.65 (2.35) & 1.89(2.13)\\

0.55& 264 & -289 ($m=0.94$ GeV) & 1.81 (2.61) & 1.99 (2.26)\\

\hline
\end{tabular} \end{center}
\end{table}

\newpage

\section{Figure captions}

\noindent Fig.1 Density dependence of the vector self energy
$\Sigma_m$ and of the scalar self energy $m^*-m$. The nuclear
matter density $\rho$ is related to its saturation value $\rho_0$.
Dotted lines: all radiation corrections are neglected. Dashed
lines: radiative corrections are included in the Leading Logarithm
Approximation (LLA). Solid lines: Corrections of the order
$\alpha_s$ are included perturbatively beyond the LLA.

 \noindent Fig.2 Density dependence of the nucleon residue $\lambda_m^2$ and of the
continuum threshold $W^2_m$. Notations are the same as in Fig.2.
\newpage

\begin{figure}[h]
\centerline{\epsfig{file=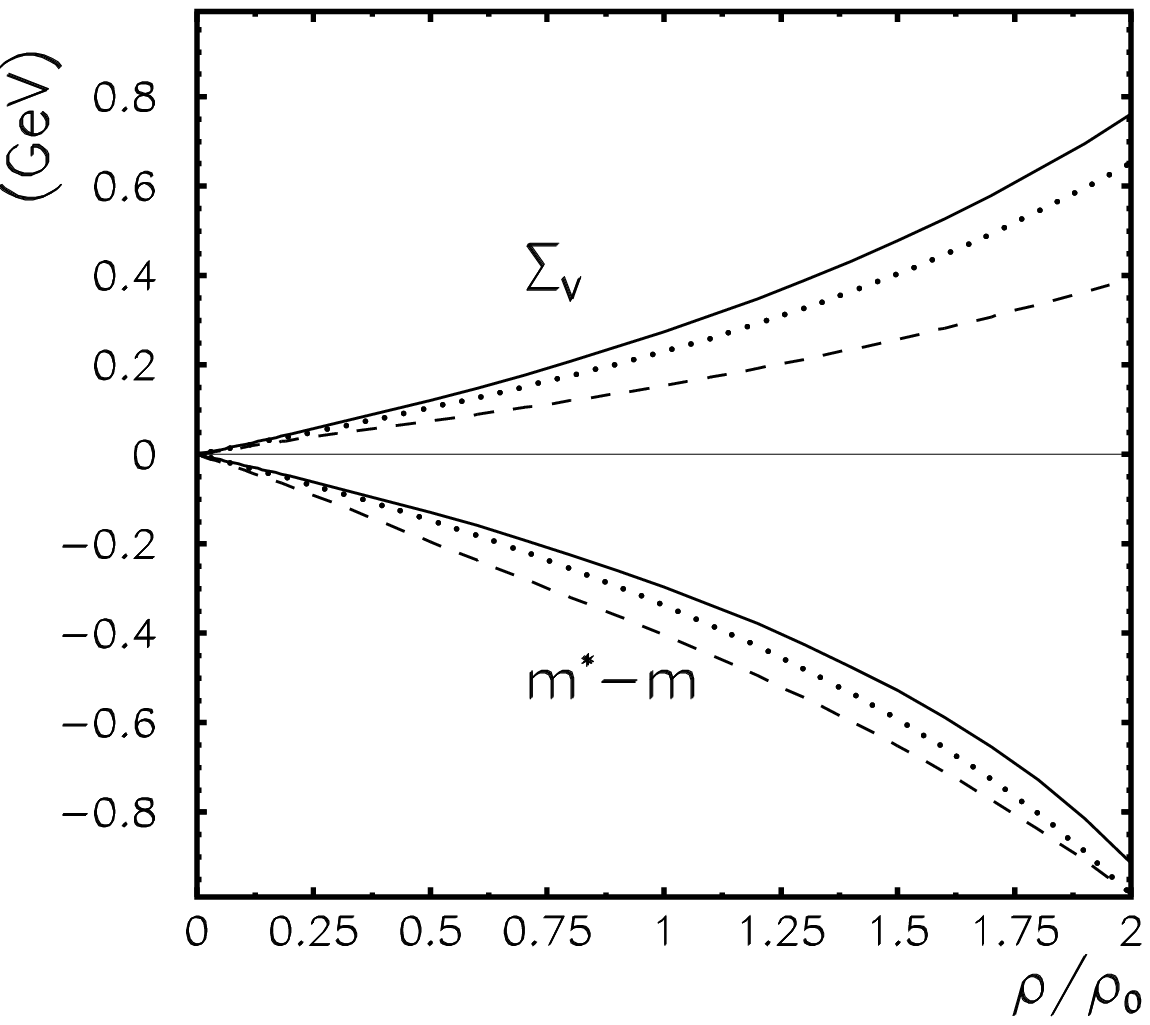,width=8.0cm}}
 \caption{}
 \end{figure}

\begin{figure}[h]
\centerline{\epsfig{file=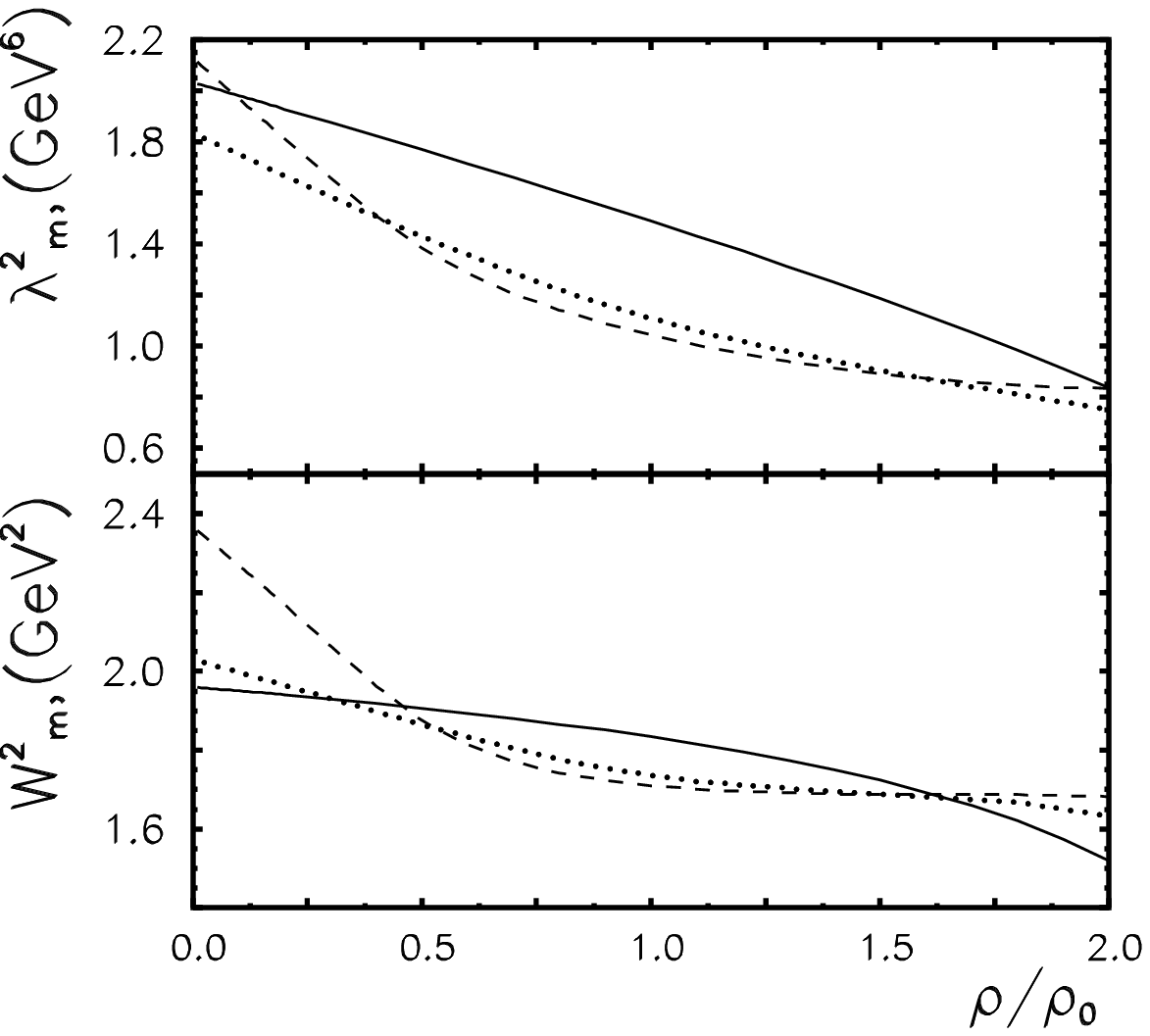,width=8.0cm}}
 \caption{}
 \end{figure}

\end{document}